\documentclass[sigconf]{acmart}
\usepackage{marvosym}
\usepackage{algorithmic}
\usepackage{graphicx}
\usepackage{textcomp}
\usepackage{colortbl}
\usepackage{xcolor}
\usepackage{diagbox}
\usepackage{enumerate}
\usepackage[figuresright]{rotating}
\usepackage{graphicx}
\usepackage{amssymb}
\usepackage{lineno}
\usepackage{multirow}
\usepackage{colortbl}
\usepackage{caption}
\usepackage{tikz,tkz-euclide,pdftexcmds}
\usepackage{booktabs} 

\usetikzlibrary{shapes,snakes}
\usetikzlibrary{positioning,arrows}
\usetikzlibrary{patterns}
\usetikzlibrary{fit}
\usetikzlibrary{backgrounds}
\setcopyright{acmcopyright}

\copyrightyear{2020}
\acmYear{2020}
\setcopyright{rightsretained}
\acmConference[SAC '20]{The 35th ACM/SIGAPP Symposium on Applied Computing}{March 30-April 3, 2020}{Brno, Czech Republic}
\acmBooktitle{The 35th ACM/SIGAPP Symposium on Applied Computing (SAC '20), March 30-April 3, 2020, Brno, Czech Republic}\acmDOI{10.1145/3341105.3374069}
\acmISBN{978-1-4503-6866-7/20/03}

\pgfdeclarelayer{foreground}

\pgfsetlayers{background,main,foreground}
\makeatletter

\pgfdeclareradialshading[tikz@ball]{my ball}{\pgfqpoint{5bp}{10bp}}{%
 color(0bp)=(tikz@ball!10!white);
 color(8bp)=(tikz@ball!65!white);
 color(12bp)=(tikz@ball);
 color(25bp)=(tikz@ball!70!black);
 color(50bp)=(black)}

\pgfdeclareradialshading[tikz@ball]{new ball}{\pgfqpoint{12bp}{12bp}}{%
 color(0cm)=(tikz@ball!10!white);
 color(0.5cm)=(tikz@ball!50!white);
 color(0.75cm)=(tikz@ball);
 color(0.9cm)=(tikz@ball!80!black);
 color(1cm)=(tikz@ball!60!black)} 

\pgfdeclareradialshading{ball shadow}{\pgfpointorigin}%
{color(0cm)=(black);
color(2mm)=(gray!70);
color(3mm)=(gray!30);
color(7mm)=(white)
} 

\tikzoption{my ball color}{\pgfutil@colorlet{tikz@ball}{#1}\def\tikz@shading{my ball}\tikz@addmode{\tikz@mode@shadetrue}}

\tikzoption{new ball color}{\pgfutil@colorlet{tikz@ball}{#1}\def\tikz@shading{new ball}\tikz@addmode{\tikz@mode@shadetrue}}

\tikzoption{ball shadow color}{\pgfutil@colorlet{tikz@ball}{#1}\def\tikz@shading{ball shadow}\tikz@addmode{\tikz@mode@shadetrue}}

\newcommand{\ball}[3][white]{%
        \begin{scope}[shift = {(#2,#3)}]
          \node[shading=ball shadow,xscale=2,yscale=0.3,circle,minimum size=5mm] 
            at (0.15,-0.0025){};
          \ifnum\pdf@strcmp{#1}{white}=\z@%
            \shade[my ball color=#1] (0.15,0.25) circle (0.2);
          \else
            \shade[new ball color=#1] (0.15,0.25) circle (0.2);
          \fi
        \end{scope}
}
\tikzset{
  pobl/.style={
    inner sep=0pt, outer sep=0pt, fill=#1,
  },
  pobl gron/.style n args={2}{
    pobl=#1, rounded corners=#2,
  },
  pics/person/.style n args={3}{
    code={
      \node (-corff) [pobl=#1, minimum width=.25*#2, minimum height=.375*#2, rotate=#3, pic actions] {};
      \node (-pen) [minimum width=.3*#2, circle, pobl=#1, outer sep=.01*#2, anchor=south, rotate=#3, pic actions] at (-corff.north) {};
      \node (-coes dde) [pobl gron={#1}{1pt}, anchor=north west, minimum width=.12125*#2, minimum height=.25*#2, rotate=#3, pic actions] at (-corff.south west) {};
      \node [pobl=#1, anchor=north, minimum width=.12125*#2, minimum height=.15*#2, rotate=#3, pic actions] at (-coes dde.north) {};
      \node (-coes chwith) [pobl gron={#1}{1pt}, anchor=north east, minimum width=.12125*#2, minimum height=.25*#2, rotate=#3, pic actions] at (-corff.south east) {};
      \node [pobl=#1, anchor=north, minimum width=.12125*#2, minimum height=.15*#2, rotate=#3, pic actions] at (-coes chwith.north) {};
      \node (-braich dde) [pobl gron={#1}{.75pt}, minimum width=.075*#2, minimum height=.325*#2, outer sep=.0064*#2, anchor=north west, rotate=#3, pic actions] at (-corff.north east)  {};
      \node [pobl=#1, minimum width=.05*#2, minimum height=.2*#2, outer sep=.0064*#2, anchor=north west, rotate=#3, pic actions] at (-corff.north east) {};
      \node (-braich chwith) [pobl gron={#1}{.75pt}, minimum width=.075*#2, minimum height=.325*#2, outer sep=.0064*#2, anchor=north east, rotate=#3, pic actions] at (-corff.north west) {};
      \node [pobl=#1, minimum width=.0375*#2, minimum height=.2*#2, outer sep=.0064*#2, anchor=north east, rotate=#3, pic actions] at (-corff.north west) {};
      \node (-fit person) [fit={(-pen.north) (-braich dde.east) (-coes chwith.south) (-braich chwith.west)}] {};
      \node (-pwy) [below=25pt of -fit person, every pin] {\tikzpictext};
      \draw [every pin edge] (-fit person) -- (-pwy);
    },
  },
}

\makeatother

\def\BibTeX{{\rm B\kern-.05em{\sc i\kern-.025em b}\kern-.08em
    T\kern-.1667em\lower.7ex\hbox{E}\kern-.125emX}}
\begin{document}
\title{An n/2 Byzantine node tolerate Blockchain Sharding approach}
\titlenote{Produces the permission block, and
  copyright information}
\author{Yibin Xu}
\orcid{0001-8490-5285}
\affiliation{%
  \institution{School of Computer Science and Informatics, Cardiff University}
  \streetaddress{Queen's Buildings, 5 The Parade}
  \city{Cardiff} 
  \state{Wales} 
  \country{UK}
  \postcode{CF24 3AA}
}
\email{work@xuyibin.top}

\author{Yangyu Huang}
\orcid{0002-0491-6871}
\affiliation{%
  \institution{School of Electronic Engineering and Automation, Guilin University of Electronic Technology}
  \city{Guilin} 
  \state{Guangxi}
  \country{China}
  \postcode{541004}
}
\email{i@hyy0591.me}
\renewcommand{\shortauthors}{Y. Xu et al.}

\begin{abstract}
Traditional Blockchain Sharding approaches can only tolerate up to n/3 of nodes being adversary because they rely on the hypergeometric distribution to make a failure (an adversary does not have n/3 of nodes globally but can manipulate the consensus of a Shard) hard to happen. The system must maintain a large Shard size (the number of nodes inside a Shard) to sustain the low failure probability so that only a small number of Shards may exist. In this paper, we present a new approach of Blockchain Sharding that can withstand up to n/2 of nodes being bad. We categorise the nodes into different classes, and every Shard has a fixed number of nodes from different classes. We prove that this design is much more secure than the traditional models (only have one class) and the Shard size can be reduced significantly. In this way, many more Shards can exist, and the transaction throughput can be largely increased. The improved Blockchain Sharding approach is promising to serve as the foundation for decentralised autonomous organisations and decentralised database.
\end{abstract}
\maketitle

\section{Introduction}
Blockchain Sharding is an approach that implements the idea of \emph{Sharding} \cite{corbett2013spanner} in blockchain to increase the transaction throughput without raising the bandwidth and processing requirements of nodes. By allowing multiple committees (Shards) running in parallel, the nodes inside every Shard solely process the data in their Shard, which leads to the system throughput increasing a lot.

Because the essence of a blockchain is in being decentralised and permissionless, and so it should allow as many devices as possible to participate in the system, the idea of Blockchain Sharding is a promising solution to solve the dilemma between increasing performance on one hand and increasing decentralisation on the other hand. Previous work on Sharding has explored various ideas (Elastico \cite{luu2016secure}, RSCoin \cite{danezis2015centrally}, OmniLedger \cite{kokoris2018omniledger}, RapidChain \cite{zamani2018rapidchain}) that can withstand up to 1/4 or 1/3 of network nodes being malicious. These approaches only support a small number of Shards in the system, or, equivalently, they require a large number of nodes in each Shard, both of which impact performance negatively.

In this paper, we propose a new Blockchain Sharding approach that can withstand up to $n/2$ of malicious nodes in the system. Compared to other methods, the probability that the malicious nodes will control a Shard is lower, and only a small number of nodes are required for every Shard to function securely. So that the communication costs inside every Shard are smaller and more Shards can exist in parallel, and that improves the transaction per second globally.

\section{Blockchain Sharding Hypothesis}
If we are inside a forest recording the time when trees fall, it is not necessarily for everyone to hear every fall of the tree to maintain the fairness of the system. The fact that a tree falls and the time when a tree falls is correct when it is recognised by most people around the tree assumed these persons have not colluded. With a sufficient number of people, if they are assigned randomly and completely distributed to subareas in the forest and are reassigned time by time to avoid the accumulation of adversary power, collusion is hard to happen (expected to occur in years). As long as the random and distributed assignment is secured, follow the principle of proportionality, taking control of a subarea requires a significant effort similar to taken the whole system when there is only one area.

In particular, this proposal is secure when (1) only people assigned to a subarea of the forest are legal to record the information about this subarea. (2) any person cannot control or predict which subarea it is about to be assigned in. (3) the assignment follows a globally recognised rule, not by the arbitrary willing of some specific group of superior people. (4) people are periodically reassigned. (5) the number of people inside every Shard is large enough.

If the above criteria are fulfilled, and with a sufficient number of honest people, one would only need to check what is the common recognised time of falling for a tree of their interest from the subarea where this tree belongs to, it is not necessary for themselves to hear the falling. In this way, people do not need to have super hearing power when the forest is dense. Instead, they only need to focus on monitoring the subarea where they are assigned to.
\begin{figure}[h]\centering
\footnotesize
\begin{tikzpicture}[scale =0.7]
\node[rectangle, fill=green!20,minimum height=2.1cm,minimum width=2.1cm,draw]at (-2.5,1.5){Forest};
\node[rectangle, fill=gray!10,minimum height=2.1cm,minimum width=2.1cm,draw]at (1.5,1.5){Forest};
\draw[thick,gray,step=.5cm] (0,0) grid (3,3);
\node[rectangle, minimum height=2.1cm,minimum width=2.1cm,draw]at (1.5,1.5){Forest};
\node [rectangle,minimum height=0.5cm,minimum width=0.5cm,fill=green!20,draw] at (1.27,0.27){ }; 
\node at (-2.9,-1.7) {\small Transitional blockchain};
\node at(1.7,-1.8){\small Blockchain Sharding} ;
\draw pic (person) [right=-60pt,pic text={\small Oversee the whole forest.}]{person={black}{25pt}{0}};
\draw pic (person) [right=35pt,pic text={\small Oversee a subarea.}]{person={black}{25pt}{0}};
\end{tikzpicture}

\label{fig:image1}
\caption{The philosophy of Blockchain Sharding}
\end{figure}
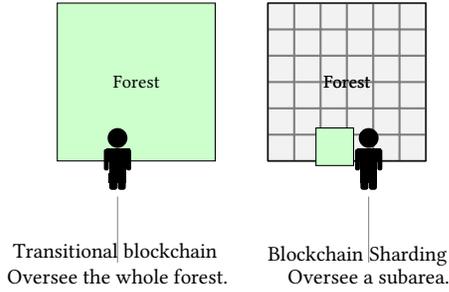
\subsection{Failure Probability}
The probability of obtaining no less than $x$ adversary nodes when randomly picking a Shard sized $m$ ($m$ is the number of nodes inside the Shard) can be calculated by the cumulative hypergeometric distribution function without replacement from a population of $n$ nodes. Let $X$ denote the random variable corresponding to the number of adversary nodes in the sampled group. The failure probability for one committee is at most
\begin{align}
Pr[X>[m/2]]&=\sum^m_{X=[m/2]}\frac{(^t_X)(^{n-t}_{m-X})}{(^n_m)}
\end{align}
which calculates the probability that no less than $X$ nodes are adversary in a group of $m$ nodes and $t$ is the number of nodes controlled by the adversary globally.

The hypergeometric distribution depends directly on the total population size (i.e., $n$). Due to $n$ can change time by time in a permissionless network (open-membership), the failure probability might be affected consequently. To maintain the desired failure probability, each Shard in RapidChain runs a consensus in pre-determined intervals (e.g. once a week), to agree on a new committee size, based on which, the committee will accept more nodes to join the committee in future epochs.

\begin{figure}[htbp]
	\begin{tabular}{c|c}
   \includegraphics[width=0.215\textwidth]{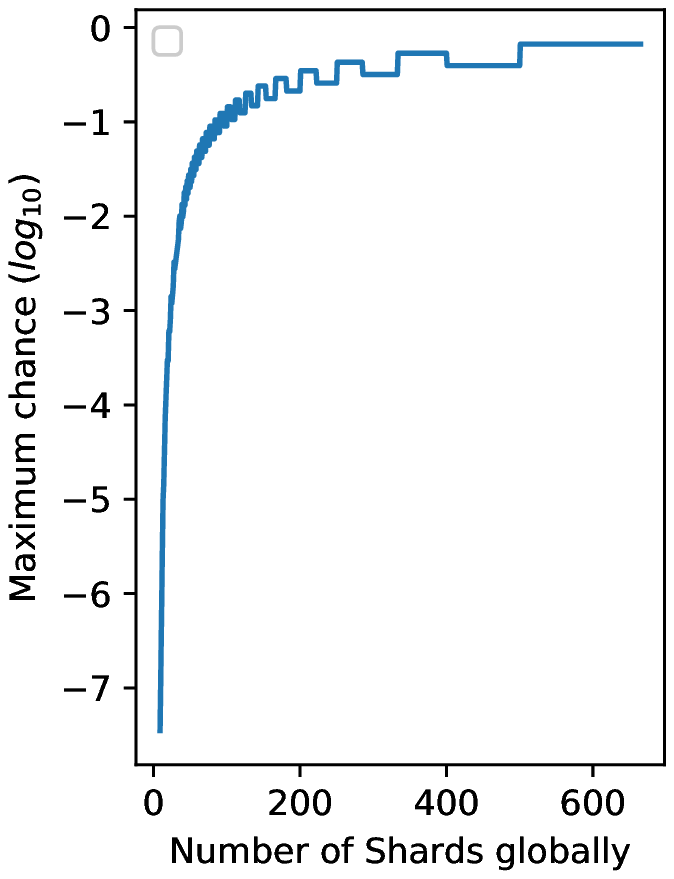}&\includegraphics[width=0.230\textwidth]{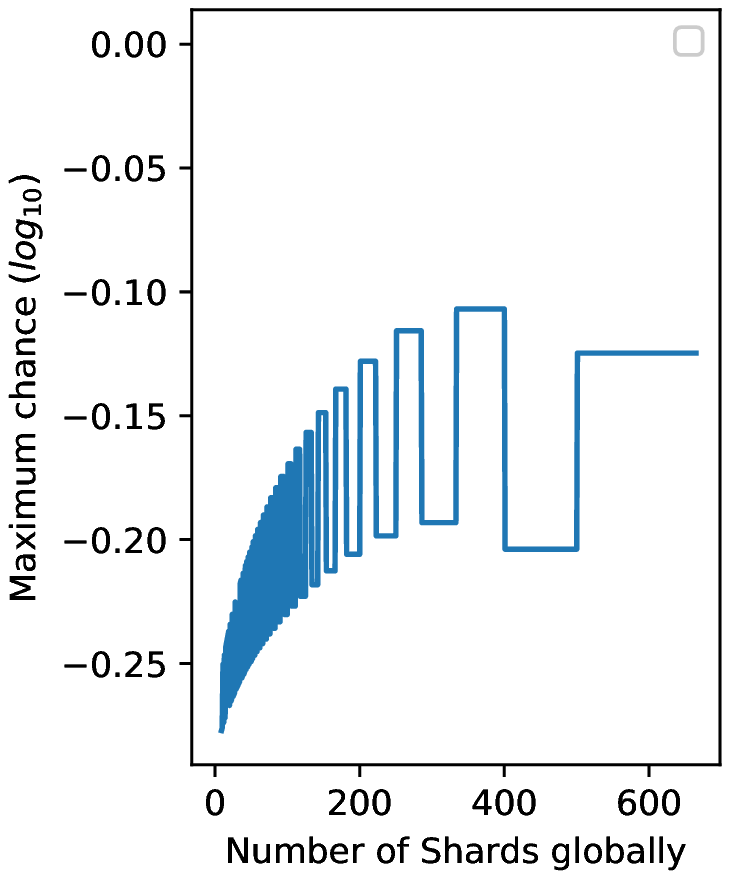}\\
    \includegraphics[width=0.215\textwidth]{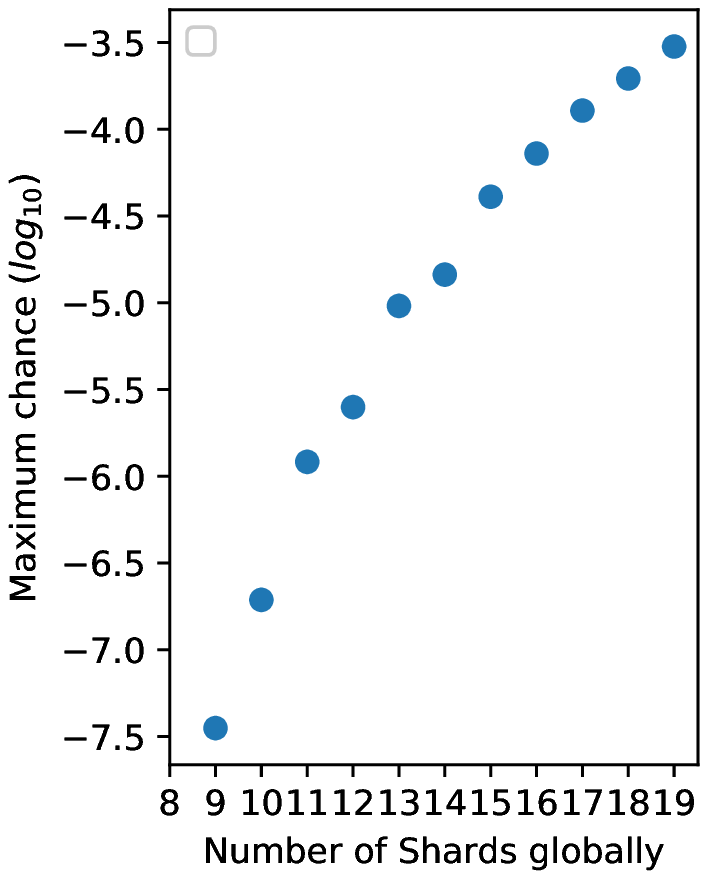}&\includegraphics[width=0.230\textwidth]{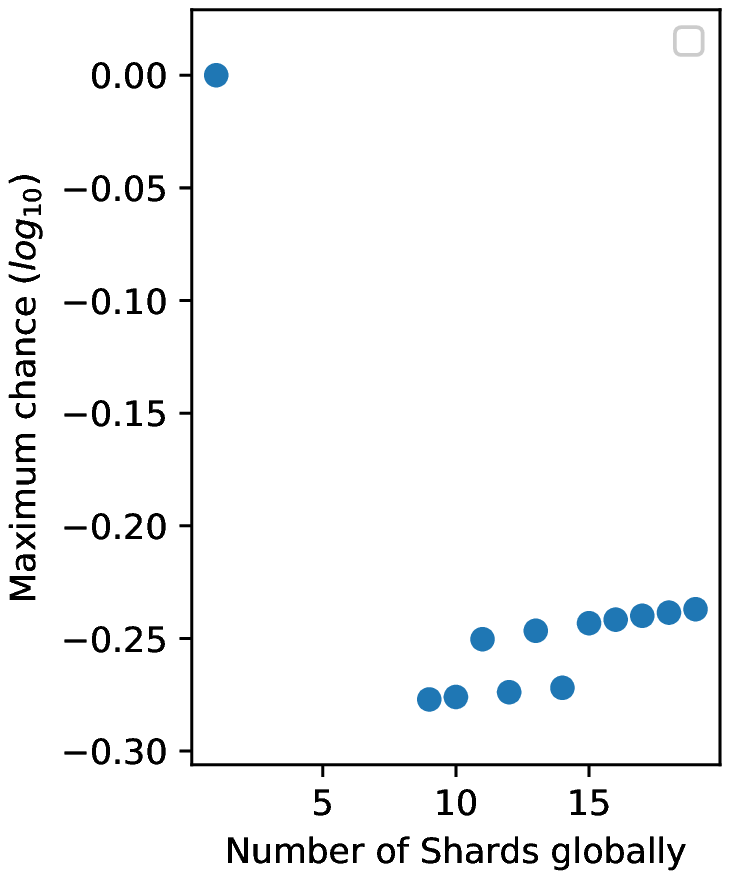}\\
		$t=n/3=666$&$t=n/2=1000$\\
\end{tabular}
	\caption{the chance to fail when $n=2000$, $t=n/3$, $t=n/2$ and $m=n/s$ where $s$ is the number of Shards;}
	\label{fig:img2}
\end{figure}

Figure \ref{fig:img2} shows the maximum probability to fail with $n=2000$ and $m=n/s$ where $s$ is the number of Shards.

As can be seen from the result, the system has a very high failure chance when the adversary taken $n/2$ of nodes, even if there are only 7 Shards. That is the main reason why all the Blockchain Sharding approaches so far are only withstanding up to $n/3$ of nodes being bad.

If every iteration lasts for $30$ minutes, then the time to secure a fail with the $10^{-6}$ failure chance with $n/3$ of nodes being bad is over $57$ years. There cannot be more than $10$ Shards when $n=2000$ and a $10^{-6}$ failure chance is maintained. The block interval (length of every synchronisation iteration) cannot be shortened; otherwise, it also reduces the time to fail. In Nakamoto blockchain, a block is published in every $10$ minute with over $1000$ transactions inside the block. Thus, the transactions embedded to the Nakamoto blockchain per hour is over $6000$. If we set the same block size for the Blockchain Sharding approaches, with $30$ minutes block interval, it can only process over $20000$ transactions per hour with $10$ Shards. However, considering all the additional designs, the lowered Byzantine fault tolerate rate, the slowed block interval and the failure in the next $57$ years, people may wonder whether a tripled performance worth all the costs.

\section{The n/2 Byzantine node tolerated Blockchain Sharding approach}
In this section, we discuss our approach in more details.
\subsection{Our Hypothesis}
Instead of recording the time of tree fallings inside a forest, imagine nodes are juries inside the courtrooms. We rule that a sentence is made when more than a predefined $T$ number of people inside the jury sized $m$ reached a consensus (the predefined $T$ must larger than $0.5 \times m$). Every jury should have $m$ people inside, and these people are from $m$ different occupations. For example, let's assume $m=5$, a jury should have five people: a teacher, a social worker, a doctor, a police officer, and a businessperson. Then, there are at least ten teachers, ten social workers, ten doctors, ten police officers, and ten businesspeople for ten juries to run in parallel. A person can choose an occupation by itself before being assigned to a jury, and it cannot change its occupation inside the jury. There is a court office which in charge of the jury membership issues. Whenever there come $m$ new people in $m$ occupations, the court office will add these people by reorganising the membership of every existing jury and then form a new jury. Jury hypothesis is distinguished from the forest hypothesis because if the adversary controls two social workers, they cannot live inside the same jury; however, they can be inside the same sub-area of the forest when recording the tree falling time. The difficulty in fulfilling the Jury hypothesis is to divide the people into $m$ occupations equally.
\subsection{Jury Membership}
A new participant needs to choose an occupation and report to the court office before the court office can assign it to a jury (Shard). When adding new people to the court system, the court office should give preference to the people of seniority (the people who reported to the office earlier but has not yet been added to the system). Assuming $m=5$ and six new people in five different occupations are waiting to be added to the system, person $A$ and person $B$ are in the same occupation. The court office would add person $A$ to the system with the other new people in other occupations if person $A$ reported to the office earlier than $B$. Person $B$ will then be put in the pending status until there comes four new people in the other four occupations.

The court office periodically publishes the number of people in every occupation who reported to the office and is waiting to be assigned to a jury. So that when a new participant decides its occupation, it will check the pending queue of every occupation and choose an unpopular occupation to get into the system quicker. If it lines in a long queue, it will need to wait until there come enough people to fill in the shorter queues. Because new participants tend to line in the shorter queue, the number of people in every occupation is automatically close to each other (tend to be equal in the long run). If people change its occupation after reporting to the court office, it will be placed to the tail of the pending queue of the new occupation. Thus, changing an occupation wastes the position in the original pending queue. 

The person, regardless if it is inside a jury or in the waiting queue, should work (generate $PoW$s) in every fixed time window. Thus the same as the other blockchain sharding models, the adversary who has half of the overall energy can only have half of the people in the system. 

Table \ref{table:r2},\ref{table:r3},\ref{table:r4},\ref{table:r5} and \ref{table:r6} show the procedure of adding people into the system. In this procedure, we rule that whenever there are at least four pending people in every occupation, adding starts. Table \ref{table:r2} shows the pending queue published by the court office at the moment one. Table \ref{table:r3} shows the pending queues when the adding conditions are meet after moment one before moment two. Table \ref{table:r4} shows the pending queue at the moment two. Table \ref{table:r5} shows the people in the court system at the moment one. Table \ref{table:r6} shows the people in the court system at the moment two where some new people are added.
\begin{table}[h!]
	\footnotesize
	\caption{The pending queues published by the court office at moment 1. Letters in red colors are pending people.}
\centering
\begin{tabular}{llllll}
Occupation &I                                                                                             & II                                                                                             & III                                              & IV                                               & V                                                \\
                        &                                                  &                                                  &                                                  &                                                  &                                                  \\
                        & \cellcolor[HTML]{FE0000}{\color[HTML]{FFFFFF} A} &                                                  &                                                  &                                                  &                                                  \\
                        & \cellcolor[HTML]{FE0000}{\color[HTML]{FFFFFF} B} &                                                  &                                                  &                                                  &                                                  \\
                        & \cellcolor[HTML]{FE0000}{\color[HTML]{FFFFFF} C} & \cellcolor[HTML]{FE0000}{\color[HTML]{FFFFFF} E} &                                                  &                                                  &                                                  \\
                        & \cellcolor[HTML]{FE0000}{\color[HTML]{FFFFFF} D} & \cellcolor[HTML]{FE0000}{\color[HTML]{FFFFFF} F} & \cellcolor[HTML]{FE0000}{\color[HTML]{FFFFFF} G} & \cellcolor[HTML]{FE0000}{\color[HTML]{FFFFFF} H} & \cellcolor[HTML]{FE0000}{\color[HTML]{FFFFFF} I} \\
Number of Pending person & 4                                                & 2                                                & 1                                                & 1                                                & 1                                               
\end{tabular}

	\label{table:r2}
\end{table}

\begin{table}[h!]
	\footnotesize
\centering
	\caption{The pending queues after moment 1 before moment 2. letters in blue color stand for the pending people who reported to the office after moment 1 before moment 2. Because the minimum length of the queues reached four (pre-defined adding parameter), the front four people of every queue should be added to the system in moment 2.}
\begin{tabular}{llllll}
Occupation                                               & I                                                & II & III                                              & IV                                               & V                                                                     \\
                                                         &                                                  &                                                  & \cellcolor[HTML]{3531FF}{\color[HTML]{FFFFFF} U} &                                                  & \cellcolor[HTML]{3531FF}{\color[HTML]{FFFFFF} V}                      \\ \cline{2-6} 
\multicolumn{1}{l|}{}                                    & \cellcolor[HTML]{FD6864}{\color[HTML]{FFFFFF} A} & \cellcolor[HTML]{6665CD}{\color[HTML]{FFFFFF} Q} & \cellcolor[HTML]{6665CD}{\color[HTML]{FFFFFF} R} & \cellcolor[HTML]{6665CD}{\color[HTML]{FFFFFF} S} & \multicolumn{1}{l|}{\cellcolor[HTML]{6665CD}{\color[HTML]{FFFFFF} T}}  \\
\multicolumn{1}{l|}{Add to court system -\textgreater{}} & \cellcolor[HTML]{FD6864}{\color[HTML]{FFFFFF} B} & \cellcolor[HTML]{6665CD}{\color[HTML]{FFFFFF} M} & \cellcolor[HTML]{6665CD}{\color[HTML]{FFFFFF} N} & \cellcolor[HTML]{6665CD}{\color[HTML]{FFFFFF} O} & \multicolumn{1}{l|}{\cellcolor[HTML]{6665CD}{\color[HTML]{FFFFFF} P}} \\
\multicolumn{1}{l|}{}                                    & \cellcolor[HTML]{FD6864}{\color[HTML]{FFFFFF} C} & \cellcolor[HTML]{FD6864}{\color[HTML]{FFFFFF} E} & \cellcolor[HTML]{6665CD}{\color[HTML]{FFFFFF} J} & \cellcolor[HTML]{6665CD}{\color[HTML]{FFFFFF} K} & \multicolumn{1}{l|}{\cellcolor[HTML]{6665CD}{\color[HTML]{FFFFFF} L}} \\
\multicolumn{1}{l|}{}                                    & \cellcolor[HTML]{FD6864}{\color[HTML]{FFFFFF} D} & \cellcolor[HTML]{FD6864}{\color[HTML]{FFFFFF} F} & \cellcolor[HTML]{FD6864}{\color[HTML]{FFFFFF} G} & \cellcolor[HTML]{FD6864}{\color[HTML]{FFFFFF} H} & \multicolumn{1}{l|}{\cellcolor[HTML]{FD6864}{\color[HTML]{FFFFFF} I}} \\ \cline{2-6} 
Number of Pending person                                  & 4                                                & 4                                                & 5                                                & 4                                                & 5                                                                    
\end{tabular}

	\label{table:r3}
\end{table}

\begin{table}[h!]
	\footnotesize
\centering
	\caption{The pending queues published by the court office at moment 2. Selected people in Table \ref{table:r3} has been assigned to the court system.}
\begin{tabular}{llllll}
Occupation              & I & II & III                      & IV & V                         \\
                        &   &   &                           &    &                           \\
                        &   &   &                           &    &                           \\
                        &   &   &                           &    &                           \\
                        &   &   &                           &    &                           \\
                        &   &   &  \cellcolor[HTML]{FE0000}{\color[HTML]{FFFFFF} U} &    &  \cellcolor[HTML]{FE0000}{\color[HTML]{FFFFFF} V} \\
Number of Pending person & 0 & 0 & 1                         & 0  & 1                        
\end{tabular}
	\label{table:r4}
\end{table}
\begin{table}[h!]
	\footnotesize
\centering
	\caption{People in the court system at moment 1. There is only one jury running.}
\begin{tabular}{llllll}
\diagbox{Ocp}{Court}     & 1                                                 &                                                 &                                                 &                                                 &                                                 \\
Occupation I   & \cellcolor[HTML]{FFFFFF}{\color[HTML]{000000} !}  & \cellcolor[HTML]{FFFFFF}{\color[HTML]{000000} } & \cellcolor[HTML]{FFFFFF}{\color[HTML]{000000} } & \cellcolor[HTML]{FFFFFF}{\color[HTML]{000000} } & \cellcolor[HTML]{FFFFFF}{\color[HTML]{000000} } \\
Occupation II  & \cellcolor[HTML]{FFFFFF}{\color[HTML]{000000} @}  & \cellcolor[HTML]{FFFFFF}{\color[HTML]{000000} } & \cellcolor[HTML]{FFFFFF}{\color[HTML]{000000} } & \cellcolor[HTML]{FFFFFF}{\color[HTML]{000000} } & \cellcolor[HTML]{FFFFFF}{\color[HTML]{000000} } \\
Occupation III & \cellcolor[HTML]{FFFFFF}{\color[HTML]{000000} \#} & \cellcolor[HTML]{FFFFFF}{\color[HTML]{000000} } & \cellcolor[HTML]{FFFFFF}{\color[HTML]{000000} } & \cellcolor[HTML]{FFFFFF}{\color[HTML]{000000} } & \cellcolor[HTML]{FFFFFF}{\color[HTML]{000000} } \\
Occupation IV  & \cellcolor[HTML]{FFFFFF}{\color[HTML]{000000} \$} & \cellcolor[HTML]{FFFFFF}{\color[HTML]{000000} } & \cellcolor[HTML]{FFFFFF}{\color[HTML]{000000} } & \cellcolor[HTML]{FFFFFF}{\color[HTML]{000000} } & \cellcolor[HTML]{FFFFFF}{\color[HTML]{000000} } \\
Occupation V   & \cellcolor[HTML]{FFFFFF}{\color[HTML]{000000} *}  & \cellcolor[HTML]{FFFFFF}{\color[HTML]{000000} } & \cellcolor[HTML]{FFFFFF}{\color[HTML]{000000} } & \cellcolor[HTML]{FFFFFF}{\color[HTML]{000000} } & \cellcolor[HTML]{FFFFFF}{\color[HTML]{000000} }
\end{tabular}
	\label{table:r5}
\end{table}
\begin{table}[h!]
	\footnotesize
	\caption{People in the court system at moment 2. All membership is adjusted with four more juries formed.}
\centering
\begin{tabular}{llllll}
\diagbox{Ocp}{Court}     & 1                                                & 2                                                & 3                                                 & 4                                                & 5                                                 \\
Occupation I   & \cellcolor[HTML]{FFFFFF}{\color[HTML]{000000} A} & \cellcolor[HTML]{FFFFFF}{\color[HTML]{000000} C} & \cellcolor[HTML]{FFFFFF}{\color[HTML]{000000} B}  & \cellcolor[HTML]{FFFFFF}{\color[HTML]{000000} D} & \cellcolor[HTML]{FFFFFF}{\color[HTML]{000000} !}  \\
Occupation II  & \cellcolor[HTML]{FFFFFF}{\color[HTML]{000000} @} & \cellcolor[HTML]{FFFFFF}{\color[HTML]{000000} Q} & \cellcolor[HTML]{FFFFFF}{\color[HTML]{000000} M}  & \cellcolor[HTML]{FFFFFF}{\color[HTML]{000000} F} & \cellcolor[HTML]{FFFFFF}{\color[HTML]{000000} E}  \\
Occupation III & \cellcolor[HTML]{FFFFFF}{\color[HTML]{000000} R} & \cellcolor[HTML]{FFFFFF}{\color[HTML]{000000} N} & \cellcolor[HTML]{FFFFFF}{\color[HTML]{000000} \#} & \cellcolor[HTML]{FFFFFF}{\color[HTML]{000000} J} & \cellcolor[HTML]{FFFFFF}{\color[HTML]{000000} G}  \\
Occupation IV  & \cellcolor[HTML]{FFFFFF}{\color[HTML]{000000} S} & \cellcolor[HTML]{FFFFFF}{\color[HTML]{000000} O} & \cellcolor[HTML]{FFFFFF}{\color[HTML]{000000} H}  & \cellcolor[HTML]{FFFFFF}{\color[HTML]{000000} K} & \cellcolor[HTML]{FFFFFF}{\color[HTML]{000000} \$} \\
Occupation V   & \cellcolor[HTML]{FFFFFF}{\color[HTML]{000000} T} & \cellcolor[HTML]{FFFFFF}{\color[HTML]{000000} *} & \cellcolor[HTML]{FFFFFF}{\color[HTML]{000000} P}  & \cellcolor[HTML]{FFFFFF}{\color[HTML]{000000} I} & \cellcolor[HTML]{FFFFFF}{\color[HTML]{000000} L} 
\end{tabular}
	\label{table:r6}
\end{table}

As can be seen from the adding procedure, if the Adversary is not in a very long queue, there is no gain for the Adversary to change the occupation once it reports to the court office. If it does so, it goes to the tail of another queue, leaving its original place to others. Then, it still needs to wait until there comes more people in other queues before it can be added to the system.

\subsection{Failure Probability}
Table \ref {fig:img3} shows a court schedule table for ten courts run in parallel with the jury sized five, and ten people in each of every occupation. In Table \ref {fig:img3}, $A$ refers to the adversary person, $H$ refers to the honest person.
\begin{table}[h!]
\footnotesize
\caption {Court Jury Schedule}
\begin{tabular}{ccccccccccc}
\diagbox{Ocp}{Court}&0&1&2&3&4&5&6&7&8&9\\
Occupation I &A&A&A&A&A&H&H&H&H&H\\
Occupation II&H&A&H&A&H&A&H&A&H&A\\
Occupation III&A&H&A&H&A&H&A&H&A&H\\
Occupation IV&H&A&H&H&A&A&H&A&H&A\\
Occupation V &H&H&H&H&A&A&A&A&H&A\\
\end{tabular}
\label{fig:img3}
\end{table}

For a $s$ number of the jury meeting to be held in parallel, there is a $s$ number of people in each occupation. Let the adversary has $A_i$ number of the person in Occupation $i$; then the chance for the adversary to secure a manipulated sentence is (assuming without loss of generality that the adversary puts all its nodes into the front $T$ occupations)
\begin{align}
 Pr[T] &= \prod_{i=1}^{T}\frac{A_i}{s}
\end{align}
where $T$ is the number of the person the adversary must take in a jury to manipulate the sentence.

To derive the maximised $Pr[T]$, we want $\prod_{i=1}^{T}A_i$ to be maximised because $s$ is the same. Let the adversary has $AD$ number of people inside the system (Court Jury Schedule), then $AD=\sum_{i=1}^m A_i$. To let the value of $\prod_{i=1}^{T}A_i$ maximise, we consider 
\begin{align}
    A_i &= \lceil(AD/T)\rceil, \quad i \in [1,AD \bmod T]\\
    A_i &= \lfloor(AD/T)\rfloor, \quad i \in (AD \bmod T, T]
\end{align}

This scenario is the maximised because given any positive integer $X$, 
\begin{align}
X*X &> (X-1)*(X+1)=X*X-1.
\end{align}
Thus, \begin{align}Pr[T]_{max}\approx(\frac{AD}{T*s})^T\end{align}

If $T=m$ (all the people in the jury should reach the same verdict when making a sentence), then \begin{align}Pr[T=m]_{max}&\approx(\frac{AD}{s*m})^m\end{align} Let $AD=\frac{s*m}{2}$ (half of the overall population) then \begin{align}Pr[T=m]_{max}\approx(\frac{1}{2})^m\end{align}

\begin{figure}[htbp]
	\begin{tabular}{l|l}
   \includegraphics[width=0.22\textwidth]{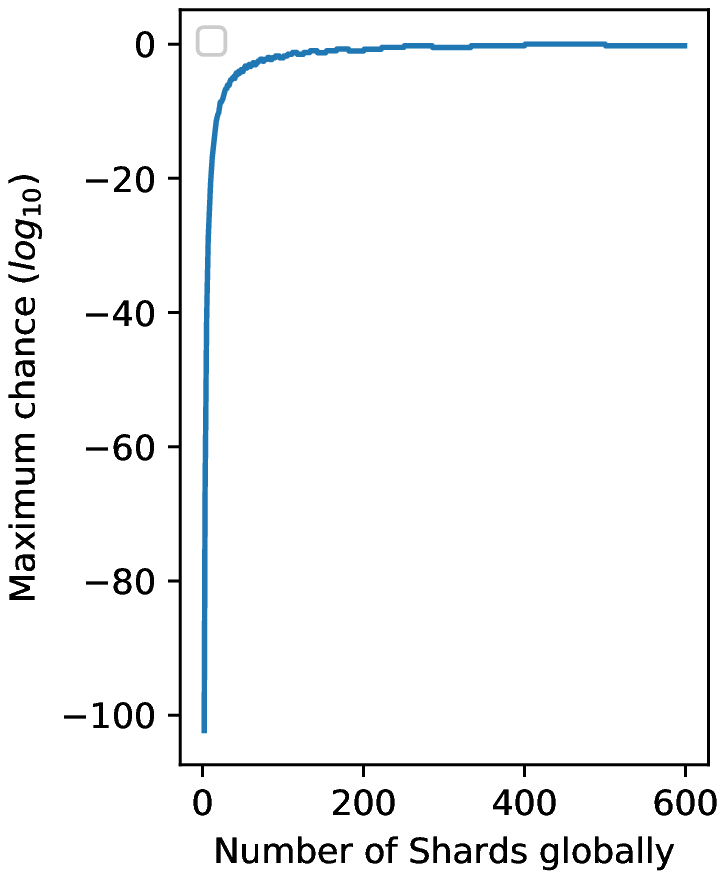}&\includegraphics[width=0.22\textwidth]{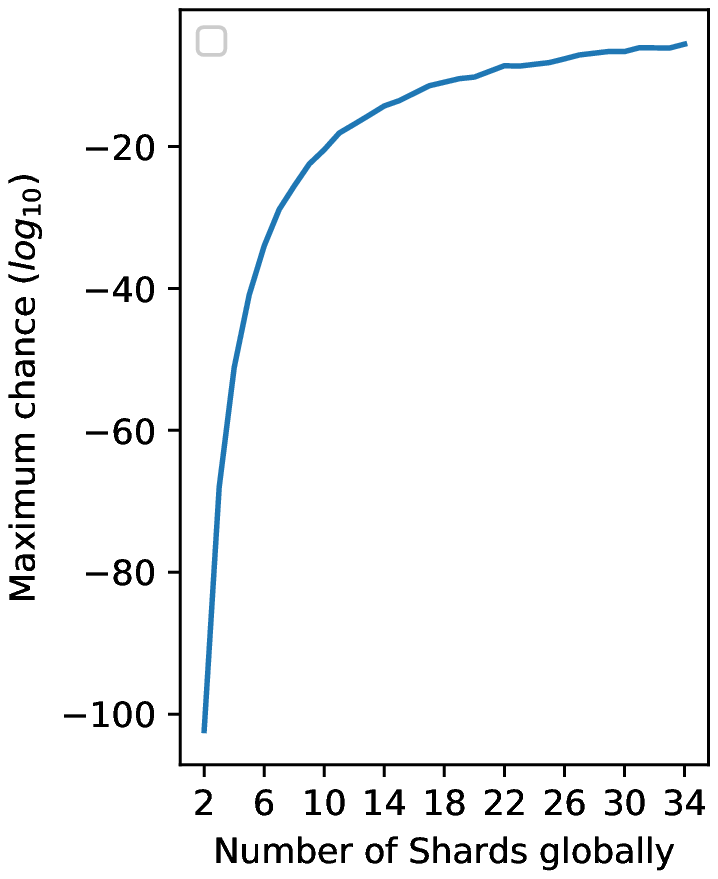}\\
		\small{$s \in [2,600], AD=1000=n/2$}&\small {$s \in [2,34], AD=1000=n/2$}\\
\end{tabular}
	\caption{the chance to fail with different $s$ when $n=2000$ and $m=n/s$ where $s$ is the number of Shards;}
	\label{fig:img4}
\end{figure}

Though the adversary cannot manipulate a sentence when it does not have $T$ people inside a Shard, it can halt a sentence to be reached when it has $m-T+1$ number of the nodes in a Shard. Then this sentence cannot be made until the next court (the group of juries are re-selected). Thus, to make the system function more smoothly, we want $T\approx[m/2]$ while meeting the security threshold (e.g. $10^{-6}$ failure chance). Figure \ref{fig:img4} shows the maximum failure chance with different $s$, $n=s*m=2000$, $T=0.7*m$ and $AD=1000$ ($1/2$ fraction of the overall population).

As can be seen from the result, when there are ten Shards and n/2 people being evil, the failure chance is below $10^{-20}$, which significantly outperformed the RapidChain at below $10^{-6}$ when it has ten Shards and only $n/3$ nodes being evil. If we set the block interval to be $30\ minutes$, then it takes over $10^{15}$ years to fail the system. If it is $10\ minutes$ (the same as Nakamoto blockchain), it still takes over $10^{14}$ years to fail. If we maintain the $10^{-6}$ failure chances at this circumstance with $T=0.7*m$, then there can be $33$ Shards at the same time.

\section{Potential usage}
It has been a long-standing question for how to open the membership in a distributed system while maintaining the performance of distributed jobs as well as the integrity and correctness of the job results \cite{ishibuchi1992distributed,friedman2003fuzzy}. How to enable nodes in the different background to participate and tolerant them go offline without notice while stabilised the system as a whole \cite{pass2017hybrid}. By increasing the Byzantine-fault-tolerant rate as well as the performance of the Blockchain Sharding approach, the improved Blockchain Sharding approach may solve these standing problems. For example, a data grid and distributed database can allow their users to be a part of the processing system. IoT devices can be governed decentralised by the transparent rule of laws \cite {psaras2018decentralised,bogner2016decentralised,mocnej2018decentralised}, in this way, waived the concern over privacy or even espionage for smart home assistants.
\section{Conclusion}
In this paper, we discussed a new Blockchain Sharding approach that maintains the system integrity when there are at most $n/2$ fraction of adversary nodes. Compared to the previous work, the required number of nodes per Shard is much lower and more Shards are allowed to exist with the same security threshold.
\bibliographystyle{unsrt}
\bibliography{sample}

\begin{thebibliography}{10}

\bibitem{corbett2013spanner}
James~C Corbett, Jeffrey Dean, Michael Epstein, Andrew Fikes, Christopher
  Frost, Jeffrey~John Furman, Sanjay Ghemawat, Andrey Gubarev, Christopher
  Heiser, Peter Hochschild, et~al.
\newblock Spanner: Google’s globally distributed database.
\newblock {\em ACM Transactions on Computer Systems (TOCS)}, 31(3):8, 2013.

\bibitem{luu2016secure}
Loi Luu, Viswesh Narayanan, Chaodong Zheng, Kunal Baweja, Seth Gilbert, and
  Prateek Saxena.
\newblock A secure sharding protocol for open blockchains.
\newblock In {\em Proceedings of the 2016 ACM SIGSAC Conference on Computer and
  Communications Security}, pages 17--30. ACM, 2016.

\bibitem{danezis2015centrally}
George Danezis and Sarah Meiklejohn.
\newblock Centrally banked cryptocurrencies.
\newblock {\em arXiv preprint arXiv:1505.06895}, 2015.

\bibitem{kokoris2018omniledger}
Eleftherios Kokoris-Kogias, Philipp Jovanovic, Linus Gasser, Nicolas Gailly,
  Ewa Syta, and Bryan Ford.
\newblock Omniledger: A secure, scale-out, decentralized ledger via sharding.
\newblock In {\em 2018 IEEE Symposium on Security and Privacy (SP)}, pages
  583--598. IEEE, 2018.

\bibitem{zamani2018rapidchain}
Mahdi Zamani, Mahnush Movahedi, and Mariana Raykova.
\newblock Rapidchain: Scaling blockchain via full sharding.
\newblock In {\em Proceedings of the 2018 ACM SIGSAC Conference on Computer and
  Communications Security}, pages 931--948. ACM, 2018.

\bibitem{ishibuchi1992distributed}
Hisao Ishibuchi, Ken Nozaki, and Hideo Tanaka.
\newblock Distributed representation of fuzzy rules and its application to
  pattern classification.
\newblock {\em Fuzzy sets and systems}, 52(1):21--32, 1992.

\bibitem{friedman2003fuzzy}
Roy Friedman.
\newblock Fuzzy group membership.
\newblock In {\em Future Directions in Distributed Computing}, pages 114--118.
  Springer, 2003.

\bibitem{pass2017hybrid}
Rafael Pass and Elaine Shi.
\newblock Hybrid consensus: Efficient consensus in the permissionless model.
\newblock In {\em 31st International Symposium on Distributed Computing (DISC
  2017)}. Schloss Dagstuhl-Leibniz-Zentrum fuer Informatik, 2017.

\bibitem{psaras2018decentralised}
Ioannis Psaras.
\newblock Decentralised edge-computing and iot through distributed trust.
\newblock In {\em Proceedings of the 16th Annual International Conference on
  Mobile Systems, Applications, and Services}, pages 505--507. ACM, 2018.

\bibitem{bogner2016decentralised}
Andreas Bogner, Mathieu Chanson, and Arne Meeuw.
\newblock A decentralised sharing app running a smart contract on the ethereum
  blockchain.
\newblock In {\em Proceedings of the 6th International Conference on the
  Internet of Things}, pages 177--178. ACM, 2016.

\bibitem{mocnej2018decentralised}
Jozef Mocnej, Winston~KG Seah, Adrian Pekar, and Iveta Zolotova.
\newblock Decentralised iot architecture for efficient resources utilisation.
\newblock {\em IFAC-PapersOnLine}, 51(6):168--173, 2018.

\end{thebibliography}
\end{document}